\documentclass[aps,prl,showpacs,twocolumn,letter,preprintnumbers]{revtex4-1}
\pdfoutput=1
\usepackage{amsmath,latexsym,amssymb,graphicx,hyperref,color,enumerate,url}
\usepackage{slashed}
\definecolor{red}{rgb}{1,0,0}
\definecolor{nicered}{rgb}{0.7,0.1,0.1}
\definecolor{nicegreen}{rgb}{0.1,0.5,0.1}
\hypersetup{colorlinks,citecolor= nicegreen,linkcolor= nicered}
\usepackage{color}
\usepackage{cancel}
\usepackage{hyperref}
\usepackage{url}

\begin{document}
\title{Top Quark Mediated Dark Matter}
%\title{Dark Matter Anomalous Couplings via Top Quark Portal}
\author{Yue Zhang}
%\email{yuezhang@ictp.it}
%\affiliation{California Institute of Technology, Pasadena, CA 91125, USA}
\affiliation{International Center for Theoretical Physics, Trieste 34014, Italy}

\begin{abstract}
We study the top quark portal dominated dark matter interactions, and its implications for the gamma ray line searches. In this picture, the dark matter interactions with photons and gluons are loop induced by the axial anomaly of the top quark current. We show there can be a natural suppression of the tree-level annihilation of dark matter, and the photon channel in turn has a substantial rate when the main annihilation proceeds into gluons. We observe a competition between the indirect detection of gamma ray line and the search with monojet plus missing energy events at LHC, and the 7 TeV data already set an upper bound of $\sim$\,$10^{-28}\,{\rm cm^3\,s^{-1}}$ on the photonic annihilation cross section. This upper limit is compatible with a thermal WIMP scenario.
\end{abstract}

\maketitle
\pagestyle{plain}

\noindent{\bfseries Introduction. }
The existence of dark matter (DM) in the universe has been established by various cosmological 
observations, yet its identity has not been uncovered. Many experiments are now going on in order to
directly or indirectly detect its trace. The indirect detection using the monochromatic cosmic gamma ray
could serve as a clear evidence for DM being from particle physics origin. 
Since the DM is electric neutral, its annihilation into two photons must happen at loop level, naively suppressed by a factor of $(\alpha/\pi)^2$ 
compared with the tree level annihilation to charged final states.
For thermal weakly interacting massive particle (WIMP) picture whose total annihilation cross section is around a picobarn,
the induced two photon annihilation rate is far below the current Fermi LAT sensitivity.

Early this year, there were several analysis~\cite{Bringmann:2012vr, Weniger:2012tx, Su:2012ft} of the 4-year Fermi data showing the positive signatures in the gamma line search from the galactic center, 
with energy 130\,GeV and a cross section $1.3\times 10^{-27}\,{\rm cm^3\,s^{-1}}$ if originating from DM annihilation~\cite{Bringmann:2012ez}.
This is only one order of magnitude below the thermal cross section.
A more recent data reprocessing performed by the Fermi-LAT collaboration shows a similar line-like feature which slightly shifts to 135\,GeV.
While the significance of this excess remains at $\sim$\,3\,$\sigma$, it seems at present one cannot exclude the uncertainties from systematics or the earth limb photon background~\cite{fermitalk}.
On the conservative side, the global analysis of Fermi data offers an upper bound on the DM to two photon annihilation cross section, which is around $10^{-27}\,{\rm cm^3\,s^{-1}}$ for photon energy 130\,GeV~\cite{Ackermann:2012qk}.

The above hint has inspired a plethora of theoretical studies~\cite{list, Chu:2012qy, Kang:2012bq} on DM models that can give an ``enhanced" gamma ray feature from annihilation.
Questions need to be addressed include: whether this signal can be reconciled with thermal DM paradigm; 
what are the correlated phenomena and in turn their constraints~\cite{correlate, Laha:2012fg, german}.
Possible answers to the former question involve the suppression or elimination of direct annihilation to light charged standard model (SM) fermions.
In many cases, additional states other than the DM itself are introduced, playing the role of co-annihilator or intermediate states (real or virtual) in the annihilation.

In this paper, we investigate a case when DM $\chi$ couples predominantly with the top quark~\cite{topwindow,Jackson:2009kg,Kang:2012bq}, and assume the couplings to other SM fermions or bosons are negligibly small throughout the discussion. 
For the bulk of this work, we consider the case when the DM is lighter than the top quark.
In this case, it cannot annihilate into both on-shell top quark and antiquark,
and we notice a fact that the three-body annihilation threshold 
$m_t + m_W + m_b \approx 256\,{\rm GeV}$,
%\begin{equation}
%m_t + m_W + m_b \approx 256\,{\rm GeV}\ ,
%\end{equation}
is only a few GeV below twice of the suspected gamma line energy in Fermi ($\sim\,$130 -- 135\,GeV), or twice of the corresponding DM mass. This indicates additional final state phase space suppression for the process $\chi\bar\chi\to t W^-\bar b$ or $\bar t W^+ b$.
Meanwhile, the DM can also annihilate into two photons via the top quark loop. The ratio between the two cross sections can be estimated as
\begin{eqnarray}
\frac{\sigma v_{\chi\bar\chi\to \gamma\gamma}}{\sigma v_{\chi\bar\chi\to tWb}} \sim 4\pi \left(\frac{\alpha}{\pi} \right)^2 \frac{m_\chi^2}{\delta m^2} \ ,
\end{eqnarray}
where the factor $4\pi$ stands for a generic ratio of two- and three-body final state phase spaces, and $\delta m \approx 2 m_\chi - m_t - m_W - m_b$ provides additional suppression when the later is being kinematically squeezed. For $m_\chi$\,$\sim$\,$130\,$GeV and $\delta m\sim$\,few GeV, the ratio can be as large as 0.1. 
It serves as one motivation to study possible enhanced gamma ray line signals with this setup.
Another advantage of the top quark portal is that the continuum gamma ray constraint turns out to be less severe~\cite{german}.

A particular feature of the top quark loop induced annihilations into $\gamma\gamma$ and $gg$ discussed here is that, they
always proceed through the axial anomaly of the top quark current.
For the mass range $m_\chi \lesssim130\,$GeV, the $\chi\bar \chi\to gg$ channel will dominate over the three-body one 
and lead to the largest ratio of photonic to the total annihilation cross section, which is $\sim\,$0.4\%.
The DM mass indicated from the Fermi data lies within this window.

The main consequences of the above picture are as follows. 
Such top quark portal being behind the Fermi gamma ray line implies a sizable effective coupling of DM to gluons. 
The total annihilation cross section exceeds that required for the thermal relic density.
This case is severely constrained by LHC measurement of missing energy plus monojet events and has been excluded using the 7\,TeV data.
On the other hand, the thermal WIMP case is still allowed by LHC data. 
The two photon annihilation cross section is about one order of magnitude smaller than that needed to explain Fermi, 
but large enough to be probed with more data and future experimental sensitivity~\cite{Bergstrom:2012vd}.
We foresee an intimate interplay between DM collider searches and indirect detections.

\medskip
\noindent{\bfseries Top Quark Portal and DM Annihilations. }
In this work, we consider the DM as a Dirac fermion and write down its effective couplings with the top quark.
In order to avoid large coupling to the lighter quarks, such as the bottom, and preserve the SM gauge symmetry, 
we choose only the coupling to the right-handed top quark. The viable operator at lowest dimensions is~\cite{other}
\begin{eqnarray}\label{portal}
\mathcal{L}_{\rm portal} = \frac{1}{\Lambda^2} \left[\bar \chi \gamma_\mu (a + \gamma_5) \chi \right] \left[ \bar t \gamma^\mu (1+\gamma_5) t \right] \ ,
\end{eqnarray}
where $\Lambda$ is the cutoff of the effective interaction and $a$ is a real number.

\begin{figure}[h]
  \centering
  \includegraphics[width=0.9\columnwidth]{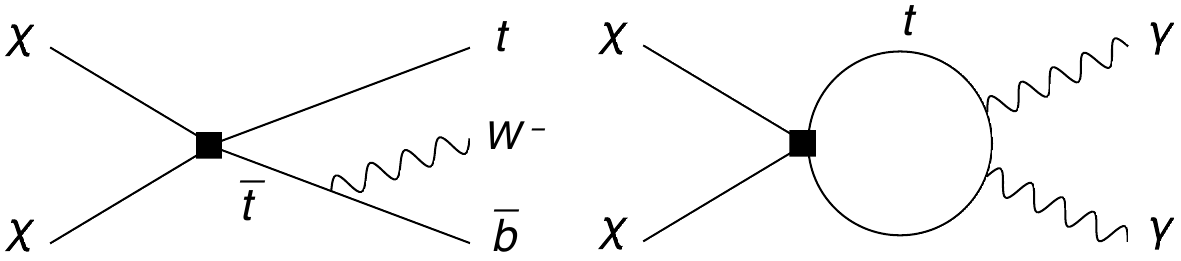}
  \caption{DM annihilation channels. Left: tree-level three-body process, kinematically suppressed in the phase space. Right: top quark loop induced annihilation into two photons.}\label{anni1}
\end{figure}

\begin{figure}[h]
  \centering
  \includegraphics[width=0.45\columnwidth]{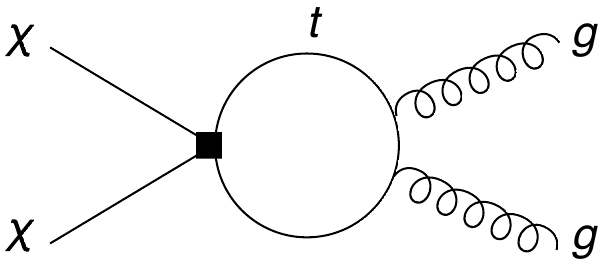}
  \caption{DM annihilate into gluons.}\label{anni2}
\end{figure}

As sketched in the introduction, the annihilation of DM has the following channels (see also Fig.~\ref{anni1} and \ref{anni2})
\begin{eqnarray}
\chi\bar\chi \to \left \{ \begin{array}{l} 
tW^-\bar b, \ \ \bar t W^+ b \\
%\gamma h\\
\gamma\gamma,\ \gamma Z\\ 
gg \ .
\end{array} \right.
\end{eqnarray}
The annihilation to $\gamma\gamma$, $\gamma Z$ and $gg$ all happen via virtual top quark loop, which we calculate in details below.

\medskip
\noindent{\it Annihilation $\chi\bar\chi\to tWb$.} 

\smallskip
\noindent For the three-body annihilation, we obtain the amplitude square
\begin{eqnarray}
&&\hspace{-0.4cm}\overline{|\mathcal{A}_{\chi\bar\chi\to tWb}|^2}  = 2 \times \frac{8 g^2}{\Lambda^4 M_W^2} \frac{m_t^2}{( s - 2\sqrt{s}\, p_3^0)^2} \nonumber \\
&&\hspace{-0.4cm}\times \left\{ (a+1)^2 (p_2 \cdot p_3) \left[ M_W^2 (p_1 \cdot p_5) + 2 (p_1 \cdot p_4) (p_4 \cdot p_5) \right]  \right. \nonumber \\
&&\hspace{-0.1cm} + (a-1)^2 (p_1 \cdot p_3) \left[ M_W^2 (p_2 \cdot p_5) + 2 (p_2 \cdot p_4) (p_4 \cdot p_5) \right]  \nonumber \\
&&\hspace{-0.1cm}\left. + (a^2-1) m_\chi^2 \left[ M_W^2 (p_3 \cdot p_5) + 2 (p_3 \cdot p_4) (p_4 \cdot p_5) \right]\right\} ,
\end{eqnarray}
where $t$, $W$ $b$ are labeled as 3, 4, 5 respectively, the pre-factor 2 takes into account of the charge-conjugation process.
%and the momentum scalar products are given in the appendix. 
Here $s\approx 4 (m_\chi + T_f)^2$ or $4m_\chi^2$ during the DM freeze out and today, respectively.
In the numerical calculation, we set the freeze out temperature $T_f\approx m_\chi/25$, in order to approximate the thermal average.

\begin{figure}[t]
\centering
\includegraphics[width=0.45\textwidth]{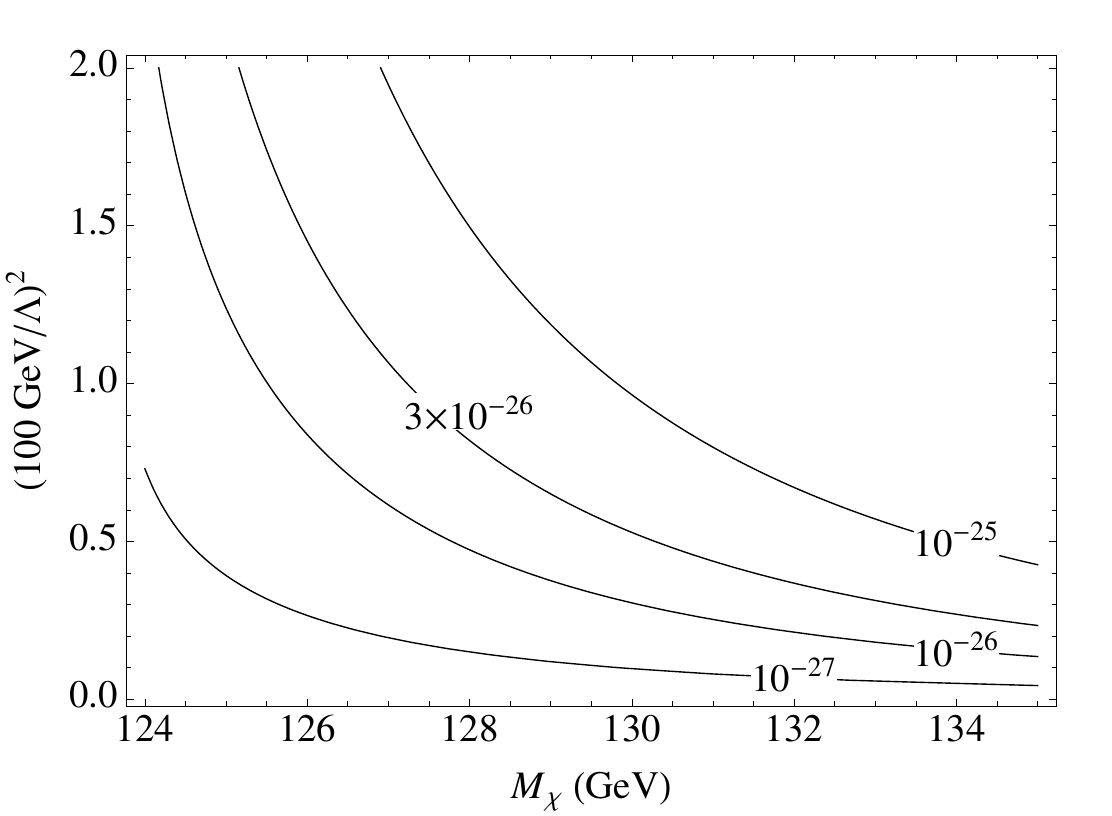}
\caption{Cross section for the three-body annihilation $\chi\bar\chi\to tWb$ during freeze-out ($T_f\approx m_\chi/25$, which corresponds to velocity $v_f\approx0.3$), in units of cm$^3/$s. Here we set $a=0$. Notice such annihilation will not happen today if $m_{\chi}\lesssim 128\,$GeV. }\label{1}
\end{figure}

We integrate over the general three-body final state phase space as given in the appendix~\cite{formcalc}.
The resulting cross section during the freeze out is shown in Fig.~\ref{1}. As one can see, the three-body annihilation can be highly suppressed by kinematics in the final state phase space, when the DM mass approaches the threshold for the $t,W, b$ final states from above.
The annihilation of DM today is more suppressed or even forbidden due to much lower velocity.

In view of this suppression, it is meaningful to examine the other loop level processes.

\medskip
\noindent{\it Annihilation $\chi\bar\chi\to \gamma\gamma$.}

\smallskip
\noindent We first examine the DM annihilation into a pair of photons, which leads to a spectacular line on top of the gamma ray spectrum.
The DM annihilation into photons can take place when the top quark lines are closed into a loop (see Fig.~\ref{anni1}).
When the tree-level annihilation is suppressed as shown above, the loop-level annihilation into photons could be more notable.

Using the interaction Eq.~(\ref{portal}), the annihilation of DM into two photons arises from the axial current $\bar t \gamma^\mu \gamma_5 t$, 
evaluated inside the triangle loop in Fig.~\ref{anni1}.
The corresponding contribution from a vector current vanishes due to the charge-conjugation invariance.
Following the generic calculation of axial anomaly~\cite{Rosenberg:1962pp, Adler:1969gk}, the finite amplitude is
\begin{eqnarray}\label{gamma}
\hspace{-1cm}\mathcal{A}_{\chi\bar\chi\to \gamma\gamma}=\frac{3 \alpha Q_t^2 m_{\chi}}{2 \pi^3 \Lambda^2} A_3(k_1, k_2) 
\bar v_\chi(p_2) \gamma_5 u_\chi(p_1) \varepsilon^{k_1k_2e_1e_2}, \hspace{-1cm}  \nonumber \\ 
%\rule{0mm}{4mm}\right. \nonumber \\
%&+&\!\!k_1^2 \left(A_3(k_1, k_2) - A_4(k_2, k_1) \right) \varepsilon^{k_2 J_\chi e_1e_2} \nonumber \\
%&+&\!\!\!\left.k_2^2 \left(A_4(k_1, k_2) -A_3(k_1, k_2) \right) \varepsilon^{k_1 J_\chi e_1e_2} \rule{0mm}{4mm}\right]\!.
\end{eqnarray}
where on-shell photon conditions have been implemented, $k_{1,2}$ are the photon momenta, and the factor 3 stands for the color factor. 
If the top quark is integrated out, the effective operator for the annihilation becomes $m_\chi \bar \chi \gamma_5 \chi F\widetilde F$.
%, whose structure is similar to that induces the $\pi^0$ decay.
The annihilation of DM to photons happens only through the axial current anomaly, and only the axial part of the dark matter current contributes.
This also agrees with the argument~\cite{GellMann:1961zz} that, the process $\nu\bar \nu\to\gamma\gamma$ vanishes if the initial neutrinos are massless.

Working in the center-of-mass frame, 
%and after imposing the on-shell conditions of final state photons, 
the form factor $A_3(k_1, k_2)$ simplifies to
\begin{eqnarray}
A_3(s) = - 16\pi^2 \int_{0}^1 dx \int_0^{1-x} dy \frac{xy}{s x y - m_t^2} \ .
\end{eqnarray}

The resulting cross section is
\begin{eqnarray}
\sigma v_{\chi\bar\chi\to \gamma\gamma} =  \frac{1}{16\pi} \left[\frac{3 \alpha Q_t^2 m_\chi}{2\pi^3 \Lambda^2} \right]^2 \left|A_3(s) \right|^2 s^2 \ .
\end{eqnarray}
As a sample value, for $m_\chi=130\,$GeV and $\Lambda=85\,$GeV, we get $\sigma v_{\chi\bar\chi\to \gamma\gamma} = 1.21\times 10^{-27}\,{\rm cm^3\,s^{-1}}$, which is close to the central value that explains the Fermi line excess. Notice a fairly low cut-off scale $\Lambda$ is required here, which indicates the higher scale physics may already be accessible.
One possible UV completion for the effective portal could be the t-channel exchange of a colored and $Z_2$ odd particle, which based on the structure of Eq.~(\ref{portal}) and upon the Fierz transformation, corresponds to $a=1$ for a vector particle, and $a=-1$ for a scalar one. 
The direct pair production and decay of such particle at LHC will result in the excess of $t\bar t + \cancel{E}_T$ events.
However, if the mass of such particle is very close to the sum of top quark and DM masses, it becomes more difficult to peel from the SM background~\cite{miha}.

\medskip
\noindent{\it Annihilation $\chi\bar\chi\to gg$.}

\smallskip
\noindent The same top quark loop also induces the annihilation of $\chi\bar\chi\to gg$, shown in Fig.~\ref{anni2}, which is a potentially dominating channel during the DM freeze out.
Its cross section can be rescaled from the two photon one, even without integrating out the top quark. 
With the electric charge $Q_t=2/3$, the ratio is fixed
\begin{eqnarray}\label{hierarchy}
\frac{\sigma v_{\chi\bar\chi\to gg}}{\sigma v_{\chi\bar\chi\to \gamma\gamma}} = \frac{9 \alpha_s^2}{8 \alpha^2} \approx 258 \ .
\end{eqnarray}

This implies an upper bound on the ratio 
\begin{eqnarray}\label{ggratio}
\frac{\sigma v_{\chi\bar\chi\to \gamma\gamma}}{\sigma v_{tot}} \lesssim 0.4\% \ ,
\end{eqnarray}
which is saturated when the DM mass is such that the annihilation into gluons dominates over the three-body one, $\sigma v_{tot}\approx \sigma v_{\chi\bar\chi\to gg}$.

\begin{figure}[t]
\centering
\includegraphics[width=0.45\textwidth]{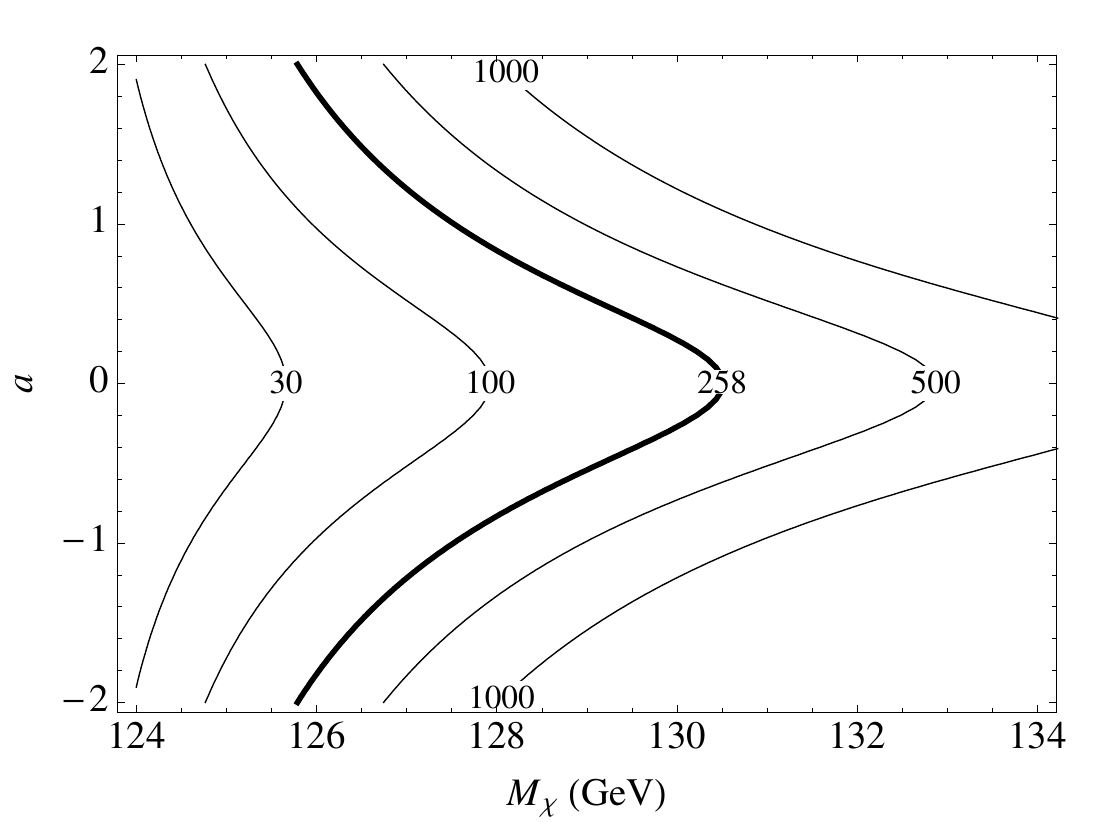}
\caption{Ratio of DM annihilation cross sections during the freeze out epoch, $\sigma v_{\chi\bar\chi\to tWb}/\sigma v_{\chi\bar\chi\to \gamma\gamma}$. 
Since the ratio $\sigma v_{\chi\bar\chi\to gg}/\sigma v_{\chi\bar\chi\to \gamma\gamma}=258$ is fixed,
we find the DM annihilation is dominated by the (anomaly loop) induced annihilation into gluons $\chi\bar\chi\to gg$ in the region to the left of the thick curve.}\label{wd}
\end{figure}

In Fig.~\ref{wd}, we compare the two potentially dominant annihilation channels, $\chi\bar\chi\to tWb$ and $\chi\bar\chi\to gg$. To the left of the thick curve, the DM annihilation during freeze out and today is controlled by the loop (axial anomaly) induced processes.
For sufficiently small $|a|$, this regime of parameter space covers the $\sim$\,130\,GeV DM mass range as hinted by Fermi.
In this case, our scenario could provide a minimal realization of the picture in~\cite{Chu:2012qy}, with a relatively smaller electric charge of the top quark though, and without introducing exotic heavy quarks.

On the other hand, if the dark matter is a thermal WIMP, from the upper bound Eq.~(\ref{ggratio}), the maximal photon annihilation is on the order of $\sim10^{-28}\,{\rm cm^3\,s^{-1}}$.

\medskip
\noindent{\it Annihilation $\chi\bar\chi\to Z\gamma$.}

\smallskip
\noindent It has been argued~\cite{Rajaraman:2012db} that the two photon annihilation of DM could be accompanied with the annihilation to $Z\gamma$, which is also the case here.
This leads to a second line in the gamma ray spectrum at 114\,(120)\,GeV for $m_\chi=130\, (135)\,$GeV.
Here $\chi\bar \chi\to Z\gamma$ is generated from the same top quark loop, but the axial current can be placed at either the $Z$ vertex or the DM portal.
We find the following ratio among different annihilation cross sections
\begin{eqnarray}
\sigma v_{Z\gamma} : \sigma v_{\gamma\gamma} \approx 0.76: 1 \ ,
\end{eqnarray}
corresponding to $m_\chi=130\,$GeV and $a=0$.
The subsequent decays of $Z$-boson into charged fermions near the center of galaxy induces synchrotron radiation, which is within the sensitivity of the galactic radio telescopes~\cite{Laha:2012fg}.

\medskip
\noindent{\bfseries Collider Implications. }
As shown in Eq.~(\ref{hierarchy}), there is a sizable ratio between the DM annihilation cross section, or the effective coupling to gluons compared to photons.
This means if the gamma ray line predicted in this scenario is large enough to be probed presently, the corresponding gluon coupling will be constrained by other experiments, {\it e.g.}, from the anti-proton flux in the cosmic ray to hadron colliders.
While the former is often subjected to larger uncertainties among the DM halo profiles and the cosmic propagation models,
the search using the monojet plus missing energy events at LHC could give a relevant and more solid constraint, which we focus on below.

In this model, the main interactions of DM relevant for the collider study are those with gluons.
Similar to Eq.~(\ref{gamma}), assuming the form factor could be approximated by a contact one in the heavy top quark limit, $A_3(k_1, k_2) \to 2\pi^2/(3m_t^2)$, the effective operator becomes $\bar \chi \gamma_5 \chi G \widetilde G$.
The parton level process is $gg\to \chi\bar\chi + jet$, with the additional jet emitted directly from the effective vertex, i.e., the top quark loop, or from initial states.

We extract the current LHC constraints on the top quark portal based on the following considerations.
The analysis in~\cite{Goodman:2010ku} shows that the two effective operators $\bar \chi \chi G G$ and $\bar \chi \gamma_5 \chi G \widetilde G$ share similar monojet constraints.
The operator ${\alpha_s}/({4 M_*^3}) \bar \chi \chi G G$ has already been constrained by the 7 TeV LHC data~\cite{ATLAS:2012ky}, which is $M_*\gtrsim 300\,$GeV for DM mass around 130\,GeV.
In the contact interaction approximation, the corresponding Wilson coefficient is 
\begin{eqnarray}
\frac{\alpha_s}{4 M_*^3} \to \frac{\alpha_s m_\chi}{3 \pi \Lambda^2 m_t^2} \ ,
\end{eqnarray}
and the ATLAS limit gets translated into
\begin{eqnarray}
\Lambda \gtrsim 200 \,{\rm GeV} \ .
\end{eqnarray}
A more precise calculation of parton level cross section keeping the explicit form factor $A_3(k_1, k_2)$ will slightly enhance the predicted signal, and leads to a relatively stronger bound on $\Lambda$. 

Both the DM annihilations to $gg$, $\gamma\gamma$ and the monojet constraint depend on a single scale $\Lambda$. 
Therefore the resulting collider constraint is closely connected to the indirect detection with gamma ray lines.
From the ATLAS result, we obtain the following implications for the top quark mediated DM scenario, which serve as the main message of this work.

\begin{itemize}
\item The lower bound on $\Lambda$ from monojet search has already exclude the parameter space that could accommodate the gamma ray line excess at Fermi. The is mainly due to the fixed and large hierarchy between the $gg$, $\gamma\gamma$ effective couplings to DM induced by the top quark loop.

\item In the window where $\chi\bar\chi\to gg$ dominates the total annihilation rate, the case of DM being a thermal candidate is still but marginally consistent with the current monojet constraint. The corresponding $\chi\bar\chi\to \gamma\gamma$ cross section is $\lesssim10^{-28}\,{\rm cm^3\,s^{-1}}$.
It will keep being tested by the upcoming data from both collider and indirect detection experiments~\cite{Bergstrom:2012vd}.
If LHC soon sees excess in the monojet events from DM, then a gamma ray line signature should also be around the corner, and vice versa.
There is an interesting competition.

\end{itemize}

For DM much heavier than 130\,GeV, and the monojet constraint gets weakened. 
Moreover, when the tree-level annihilation dominates the freeze out, one needs to take higher value of $\Lambda$ for it to be thermal.
In this case, the contact top quark portal may be probed by other channels such as $t\bar t + \cancel{E}_T$, where a pair of DM is produced in together with top-anti-top pair~\cite{topwindow}.

\medskip
\noindent{\bfseries Direct Detection. } 
In this section, we briefly sketch the direct detection of DM from the above top quark portal.
At the energy scale of DM-nucleon scattering, the top quark is heavy and can be integrated out. 

First, the anomaly loop induces a set of non-conventional effective gluonic operators for the direct detections.
The effective interactions are,
\begin{eqnarray}\label{Leff}
\hspace{-0.4cm}\mathcal{L}_{\rm eff}\!=\!\frac{\alpha_s}{3 \pi \Lambda^2 m_t^2}\!\left[ m_\chi \bar\chi i \gamma_5 \chi G_{\mu\nu}\widetilde G^{\mu\nu} 
\!+\! \bar \chi \gamma^\mu \gamma_5 \chi (D_\alpha G^{\alpha\rho}) \widetilde G_{\rho\mu}\right], \hspace{-0.7cm}  \nonumber \\
\end{eqnarray}
where $\widetilde G^{\mu\nu} = \frac{1}{2} \epsilon^{\mu\nu\rho\sigma} G_{\rho\sigma}$. In the second operator, the derivative on $G^{\alpha\rho}$ has been promoted to a covariant one in respect of gauge invariance. It vanishes when considering DM annihilating into gluons due to the on-shell condition, but will contribute here to the matrix element between nucleons. Both operators lead to spin-dependent scatterings.

The nucleon matrix element for the first gluonic operator is
\begin{eqnarray}
\langle N| G_{\mu\nu}\widetilde G^{\mu\nu} |N \rangle = \frac{32\pi^2}{3} g_A^{(0)} m_N \bar u_N i\gamma_5 u_N \ ,
\end{eqnarray}
where $g_A^{(0)} \approx 0.36$~\cite{Diakonov:1995qy}, and $N=p, n$ stands for proton and neutron, respectively. The resulting amplitude $(\bar u_\chi \gamma_5 u_\chi)(\bar u_N \gamma_5 u_N)$, 
and in turn the differential cross section $d \sigma/d E_R$, is suppressed by the momentum transfer of the DM-nucleon scattering, which is apparent after the non-relativistic expansion.
For the choice of parameter $\Lambda$ that can account for the monochromatic photon cross section within the Fermi LAT sensitivity, the corresponding spin-dependent cross section is found to be less than $10^{-20}\,$pb, far below the recent direct detection~\cite{Behnke:2012ys, Garny:2012it} and neutrino telescope~\cite{Tanaka:2011uf} bounds.

The second operator in Eq.~(\ref{Leff}) is equivalent to, by using the equation of motion,
\begin{eqnarray}
\langle N| (D_\alpha G^{\alpha\rho}) \widetilde G_{\rho\mu} |N \rangle &=& \sum_f \langle N| g_3 \bar q_f \widetilde G_{\mu\rho} \gamma^\rho q_f |N \rangle \nonumber \\
&=& \sum_f 2 f^N_{2f} m_N^2 s_\mu \ .
\end{eqnarray}
It is a spin-dependent twist-4 operator, whose matrix element is proportional to the nucleon polarization $s_\mu$~\cite{Ji:1997gs}.
As a rough estimate of the coefficients $f_{2f}^N$, assuming up/down quark flavor dominance inside a proton/neutron state, one can infer the sum of coefficients from the one ($f_2^p=\sum_f e_q^2 f_{2f}^p$) quoted in~\cite{Ji:1997gs} and estimate
\begin{eqnarray}
f^p_{2u}+f^p_{2d}+f^p_{2s} \approx (9/4) f_2^p \sim 0.1 \ .
\end{eqnarray}
%Its matrix element is similar to that of an axial current~\cite{Ji:1993sv, Belanger:2008sj}
%\begin{eqnarray}
%\langle N| \bar q_f \gamma_\mu \gamma_5 q_f |N \rangle = 2 a^N_{0f} m_N^2 s_\mu \ .
%\end{eqnarray}
Therefore, such operator also leads to a spin-dependent DM-nucleon interaction, with effective operators at nucleon level, $2 \xi_N (\bar \chi \gamma^\mu \gamma_5 \chi)(\bar N s_\mu N) $. The coefficient is
\begin{eqnarray}
\xi_N= \frac{\alpha_s}{3 \pi \Lambda^2 m_t^2}\,(f^N_{2u} + f^N_{2d} + f^N_{2s}) \, m_N^2 \ .
\end{eqnarray}
%Numerically, $\xi_p = ...$ and $\xi_n = ...$. 
The resulting spin-dependent direct detection cross section on a proton is
\begin{eqnarray}
\sigma^{\rm SD} _{\chi p} = \frac{6}{\pi} \mu^2 \xi_p^2 \ ,
\end{eqnarray}
where $\mu$ is the reduced mass $\mu=m_p m_\chi /(m_p+m_\chi)$.
Using the same values of parameters as above, the corresponding spin-dependent cross section on a proton is less than $10^{-10}\,$pb, also well below the current experimental limits. 

The reason for this suppression can also be understood. This effective operator is dimension eight, suppressed by $1/(\Lambda^2 m_t^2)$. The nucleon matrix element of the gluonic part of operator is typically of order $m_N^2$. This leads to an additional suppression of $(m_N/\Lambda)^4$ in the final cross section, compared to the usual WIMP case. 
%We point out that the smallness of the direct detection cross sections is mainly due to the contact top quark portal to DM chosen to be discussed here.
A possible way to enhance such anomalous top quark current induced DM scattering to probeable level could be trading the effective scale suppression $1/\Lambda^2$ for sub-GeV force mediators.
However, this would require further model building and the existence of new light states could in turn modify the above DM freeze out processes and detections. We leave a systematic study of it for a future work.

It turns out the most promising direct detection of such DM from the top quark portal is the exchange of a single photon in the t-channel~\cite{Agrawal:2011ze}, which is spin independent. Although this operator is suppressed by the small electromagnetic coupling, the photon exchange yields a dimension 6 operator, therefore relatively enhanced by a larger nucleon matrix element, compared to the gluonic operators.

Without specifying the UV complete theory, we could estimate the corresponding cross section averaged per nucleon as
\begin{eqnarray}
\sigma_{\chi p}^{\rm SI} \sim \frac{\mu^2 Z^2}{\pi A^2} \left( \frac{3 \alpha Q_t}{2\pi \Lambda^2} \right)^2 \ ,
\end{eqnarray}
in the case of $a=-1$~\cite{Agrawal:2011ze}.
The cross section is a few $10^{-45}\,{\rm cm}^2$ for the xenon target and $\Lambda\sim200\,$GeV (following the LHC monojet constraint).
This is close to the latest Xenon 100 limit~\cite{Aprile:2012nq}, and the parameter space will continue to be tested with more data, and the future Xenon 1T~\cite{Aprile:2012zx} and LUX~\cite{Akerib:2012ys} experiments.

\medskip
\noindent{\bfseries Conclusion and Outlook. }
To summarize, we have studied a scenario where the DM interacts with the SM sector predominantly via the top quark portal. 
This picture is partly motivated by the recent hint of gamma ray line excess in the Fermi data. 
The tree level annihilation with one top quark off-shell can be kinematically suppressed, for the DM mass near 130\,GeV.
There is a window where the DM annihilation is dominated by axial current anomaly induced coupling to gluons (photons).
In this window, the annihilation into photons is at a percent level of the total cross section. 
We show the current 7\,TeV LHC monojet data is already in tension with the gamma ray hint at Fermi in this picture.
In contrast, a thermal WIMP is marginally allowed by the LHC data, 
and the corresponding monochromatic photon flux is within the sensitivity of current and planned indirect detection experiments.

It will be interesting to see the future interplay between the DM searches at colliders and indirect detections.

\medskip
\noindent{\bfseries Acknowledgement. }
I thank Shao-Long Chen, Basudeb Dasgupta, Zhi-Hui Guo, Miha Nemev\v{s}ek, Goran Senjanovi\'c, Michel Tytgat and Gabrijela Zaharijas for useful conversations and reading the manuscript, and Z.~Chacko for pointing out the importance of single photon exchange process in the direct detection.

\medskip
\noindent{\bfseries Appendix: three-body annihilation. }
Here we give the general final-state phase space integration for the three-body annihilation cross section $\chi(1)\bar\chi(2)\to t(3) W(4) b(5)$, following the notation in~\cite{formcalc},
\begin{eqnarray}
\hspace{-0.6cm}\sigma\! =\!\! \int_{m_b}^{(p_5^0)_{\rm max}}\!\! d p_5^0\! \int_{(p_3^0)_{\rm min}}^{(p_3^0)_{\rm max}}\!\! d p_3^0\! \int_{-1}^1\!\! d \cos\theta \int_0^{2\pi} d \eta \frac{\overline{|\mathcal{A}|^2}}{8 (2\pi)^4 \Phi}\ , \hspace{-0.6cm} \nonumber\\
\end{eqnarray}
where $\theta$, $\eta$ is the angle between ${\bf p_5}$ and the $\hat z$, $\hat y$ axis, respectively,
and $\Phi = 4 \sqrt{s}\, |{\bf k}_1|$. 
All the the relevant momentum products in the amplitude square can be expressed as functions of independent variables $p_3^0$, $p_5^0$, $\theta$, $\eta$.
The integral limits are $(p_5^0)_{\rm max}= \sqrt{s}/2 - [(m_t+M_W)^2 - m_b^2]/(2\sqrt{s})$, and $(p_3^0)^{\rm max}_{\rm min}=
\frac{1}{2\tau}\left[ \sigma (\tau + m_+m_- ) \pm \sqrt{(p^0_5)^2-m_b^2} \sqrt{(\tau - m_+^2)(\tau - m_-^2)} \right]$, with $\sigma = \sqrt{s} - p_5^0$, 
$\tau = \sigma^2 - (p^0_5)^2+m_b^2$, $m_\pm=m_t \pm M_W$.

%\begin{eqnarray}
%p_3 \cdot p_4 &=& \frac{1}{2} (s + m_b^2 - m_t^2 - M_W^2) - \sqrt{s}\, p_5^0, \nonumber \\
%p_4 \cdot p_5 &=& \frac{1}{2} (s + m_t^2 - m_b^2 - M_W^2) - \sqrt{s}\, p_3^0, \nonumber \\
%p_3 \cdot p_4 &=& \sqrt{s}\, (p_3^0+p_5^0) -  \frac{1}{2} (s + m_b^2 + M_W^2 - m_t^2),  \nonumber \\
%p_1 \cdot p_3 &=& \frac{\sqrt{s}}{2} p_3^0 - \sqrt{\frac{s}{4} - m_\chi^2} \sqrt{(p_3^0)^2 - m_t^2} \cos\theta \cos\xi, \nonumber \\
%p_1 \cdot p_5 &=& \frac{\sqrt{s}}{2} p_5^0 - \sqrt{\frac{s}{4} - m_\chi^2} \sqrt{(p_5^0)^2 - m_b^2} \cos\theta, \nonumber \\
%p_1 \cdot p_4 &=& \frac{s}{2} - p_1 \cdot p_3 - p_1 \cdot p_5, \nonumber \\
%p_2 \cdot p_3 &=& \frac{\sqrt{s}}{2} p_3^0 + \sqrt{\frac{s}{4} - m_\chi^2} \sqrt{(p_3^0)^2 - m_t^2} \cos\theta \cos\xi, \nonumber \\
%p_2 \cdot p_5 &=& \frac{\sqrt{s}}{2} p_5^0 + \sqrt{\frac{s}{4} - m_\chi^2} \sqrt{(p_5^0)^2 - m_b^2} \cos\theta, \nonumber \\
%p_2 \cdot p_4 &=& \frac{s}{2} - p_2 \cdot p_3 - p_2 \cdot p_5, \nonumber \\
%\cos\xi &=& \frac{s+m_t^2+m_b^2+2 p_3^0 p_5^0 - 2 \sqrt{s}\,(p_3^0+p_5^0)}{2\sqrt{(p_5^0)^2 - m_b^2} \sqrt{(p_3^0)^2 - m_t^2}} \ . \nonumber
%\end{eqnarray}


\begin{thebibliography}{99}
\bibitem{other}
We do not consider the operator like $\bar \chi \chi \bar t_L t_R \langle H\rangle$, which requires a counterterm for $\bar \chi \chi \bar b_L b_R \langle H\rangle$ through renormalization and may upset our assumption.
\end{thebibliography}
\end{document}